\documentclass[11pt]{article}

\usepackage[paper=a4paper,dvips,top=2.0cm,left=2.2cm,right=2.2cm,bottom=2.6cm,footskip=1.3cm,voffset=0.8cm]{geometry}
\usepackage{amsfonts,amsmath,amsthm,amssymb}
\usepackage{mathrsfs}
\numberwithin{equation}{section}  
\usepackage[svgnames]{xcolor}
\usepackage{fix-cm}
\usepackage{sectsty}
\usepackage{fancyhdr}
\usepackage{graphicx}
\usepackage{url}
\usepackage{enumerate}
\usepackage{paralist}
\usepackage{multicol}
\usepackage{color}  
\usepackage{float}  
\usepackage{epsfig}
\usepackage{enumitem}
\usepackage{hyperref}   
\hypersetup{colorlinks=true,linkcolor=beamer@PRD, citecolor=beamer@PRD}
\usepackage{authblk}  
\usepackage{cite}  

\newcommand\myref[1]{\textcolor{beamer@PRD}{(}\ref{#1}\textcolor{beamer@PRD}{)}}

\usepackage{soul}  
\newcommand{\lambdabar}{{\mkern0.75mu\mathchar '26\mkern -9.75mu\lambda}} 
\definecolor{beamer@blue}{RGB}{0,0,255}
\definecolor{beamer@mediumblue}{RGB}{0,0,190}
\definecolor{beamer@midnightblue}{RGB}{25,25,112}
\definecolor{beamer@navy}{RGB}{0,0,128}
\definecolor{beamer@darkblue}{RGB}{0,0,139}
\definecolor{beamer@purple}{RGB}{128,0,128}
\definecolor{beamer@levander}{RGB}{100.,149.,237.}
\definecolor{beamer@PRD}{RGB}{46,48,146}
\definecolor{beamer@green}{RGB}{0,128,0}
\definecolor{beamer@darkgreen}{RGB}{0,100,0}
\definecolor{beamer@olive}{RGB}{128,128,0}
\definecolor{beamer@darkolivegreen}{RGB}{85,107,47}
\definecolor{beamer@gray}{RGB}{190,190,190}
\definecolor{beamer@ivry}{RGB}{220,220,220}
\definecolor{beamer@new}{RGB}{40,120,50}
\definecolor{shadecolor}{RGB}{220,220,220}
\definecolor{beamer@darkslategray}{RGB}{47,79,79}
\definecolor{beamer@chocolate}{RGB}{210,105,30}
\definecolor{beamer@brown}{RGB}{165,42,42}
\definecolor{beamer@orangered}{RGB}{255,69,0}
\definecolor{beamer@maroon}{RGB}{255,0,0}

\begin{document}


\title{\textbf{Path integral for non-paraxial optics}}
\author{Maria Chiara Braidotti$^{1,2} $\footnote{{ Now at School of Physics and Astronomy, University of Glasgow, Glasgow, UK}}, Claudio Conti$^{2, 3}$, Mir Faizal$^{4, 5}$, Sanjib Dey$^6$, \\  \vspace{-0.3cm}Lina Alasfar$^7$\footnote{{ Now at Max-Planck-Institute for Nuclear Physics.  Saupfercheckweg 1, 69117 Heidelberg
			Germany  }},  Salwa Alsaleh$^8$,  and Amani Ashour$^9$}
\date{\small{\textit{$^1$Department of Physical and Chemical Sciences, University of L'Aquila, \\ 
Via Vetoio 10, I-67010 L'Aquila, Italy \\
$^2$Institute for Complex Systems, National Research Council (ISC-CNR), \\
Via dei Taurini 19, 00185 Rome, Italy \\
$^3$Department of Physics, University Sapienza, Piazzale Aldo Moro 5, 00185 Rome, Italy\\
$^4$Irving K. Barber School of Arts and Sciences, University of British Columbia-Okanagan, \\ 
3333 University Way, Kelowna, British Columbia V1V 1V7, Canada \\
$^5$Department of Physics and Astronomy, University of Lethbridge, Lethbridge, Alberta, T1K 3M4, Canada\\
$^6$Department of Physical Sciences, Indian Institute of Science Education and Research Mohali, \\
Sector 81, SAS Nagar, Manauli 140306, India \\
$^7$ Universit\'{e} Clermont Auvergne,\\4, Avenue Blaise Pascal 63178 Aubière Cedex, France \\
$^8$Department of Physics and Astronomy, College of Science, King Saud Universty,\\
 Riyadh 11451, Saudi Arabia\\
$^9$Mathematics Department, Faculty of Science, Damascus university, Damascus, Syria}}}
\maketitle
    	
\begin{abstract}
In this paper, we have constructed  the Feynman path integral method for   
non-paraxial optics. 
This is done by using the mathematical analogy between a non-paraxial  
optical system and the generalized Schr\"odinger equation
deformed by the existence a minimal measurable length. 
Using this analogy, we  investigated the consequences  of   a minimal length   in  this optical system. 
 This path integral has been used   to obtain  
  instanton solution   for such a optical systems. Moreover,  the Berry phase of this optical 
 system has been investigated. These results may disclose a new way to use the path integral approach in optics. Furthermore, 
 as such system with an intrinsic minimal length have been studied in quantum gravity, the ultra-focused optical 
 pluses can be used as an optical analog of quantum gravity.  
\end{abstract}	 
\section{Introduction} \label{sec1}
\addtolength{\voffset}{-0.8cm} 
\addtolength{\footskip}{0.8cm} 
Non-paraxial optics is an interesting branch of research, 
which is a generalization of the standard paraxial optics \cite{Ciattoni:00}. 
Under this framework, one can describe light propagation when the waist of the laser beam is comparable to 
the diffraction length \cite{Ciattoni:00}, or when the spectral width of a pulse is much smaller than 
the pulse central frequency \cite{AgrawalBook}. 
These nonlinear techniques are of great use in recent days, for instance, these kind of beams have been utilized for 
the generation of sub-wavelength anti-diffracting beams in order to obtain super-resolved microscopy 
\cite{Klar1999,Delre2015}. Standard optical theories cannot model these systems. Rather, in some recent studies it 
came out that it is possible to study non-paraxial optical systems using a generalized uncertainty principle (GUP). 
More precisely, it has been shown in \cite{Conti2014,Braidotti2016} that the propagation equation of the 
first order for non-paraxial beams and ultra-short pulses is equivalent to 
a GUP deformed free particle Schr\"odinger equation.
This mathematical analogy is based on the fact that such 
systems have an intrinsic minimal measurable length associated with them. 
Such a  minimal measurable length also occurs in quantum gravitational systems
\cite{z4,KMM,Brau,Das,Gomes,Bagchi,Quesne,Pedram,Hossenfelder}, and so the non-paraxial optics can be 
studied as a optical analog of quantum gravitational system. So, such optical 
systems can considered as quantum gravity analog in optics. We would like to point out that the study 
of quantum gravity analogs in low energy systems has become important, as for example, many quantum gravity analogs have been proposed in acoustics \cite{unruh1981},
Bose-Einstein condensates \cite{Garay_1,Steinhauer2014_1,Steinhauer:2015saa,deNova:2018rld} and optics
\cite{Marino2008_1,FaccioBook,ng,phyrevb}. 

It is worth noting that many quantum gravity theories account for a minimal length, it occurs in 
string theory, where  the string length is considered to be the fundamental length 
scale, and it is not possible to probe the space-time below this scale
\cite{Gross,a,Yoneya,Konishi}.  Such a minimal measurable length 
scale even occurs in loop quantum gravity \cite{Rovelli,z1}. Furthermore,
black hole physics argues that any theory of quantum gravity should be composed of an intrinsic
minimal measurable length scale \cite{z4,z5}. Nevertheless, this minimal length is not a manifestation
of the Heisenberg uncertainty principle, as within this framework, the position of a particle can be 
measured with a arbitrary high accuracy, if momentum is not measured. But, in order to establish
the notion of a minimal length, a modification of the Heisenberg uncertainty principle 
has been appealed, which is commonly known as the GUP
in the literature, and its explicit form is given by
\begin{equation}\label{GUP}
\Delta X\Delta P\geq\frac{1}{2}\Big|\big\langle [X,P]\big\rangle\Big|.
\end{equation}
Here $X$ and $P$ represent the observables corresponding to the 
generalized model and, in general, the commutation relation between the observables involves 
a higher order term in momentum in order to lead to a system exhibiting minimal length. Nevertheless, 
the minimal measurable length scale, that emerges out in most of the cases, is of the order of Planck length and hence it is not possible to test all the theories that accounts for it directly as current technologies 
involving scattering experiments are not able to reach this scale. Therefore, many alternative schemes 
have been proposed which may provide deeper insights to quantum gravity:
see for instance \cite{Pikovski,Marin,Bawj,Dey_NPB}. So, such model with minimal length have been thoroughly studied 
in quantum gravity, and is mathematically and theoretically well established. 

Now as these models with a minimal length in 
quantum gravity mathematically resemble a non-paraxial optical system  \cite{Conti2014,Braidotti2016}, and 
  the path integral \cite{Feynman1948} for the Schr\"odinger equation deformed by a minimal length has been obtained 
  \cite{Das1, Das2}, 
we use it to obtain the path integral for non-paraxial optical systems. 
In this paper, we use this formalism developed to the path integral kernel of the 
GUP deformed Schr\"odinger equation to analyze a non-paraxial optical system. 
We analyze several optical effects using this path integral, especially we show that it enhances 
the wave-particle duality property of light as well as it imposes a spectral limitation. 
Furthermore, the approach allows us to find the upper bound of the momentum corresponding to the minimal length scale. 
We  have also computed the instantons and  Berry  phase corresponding to the non-paraxial optical system. 

 
\section{ Non-paraxial Optics as a Quantum Gravity Analog}\label{sec2}
\lhead{Path integral for non-paraxial optics}
\chead{}
\rhead{}
\addtolength{\voffset}{0.8cm} 
\addtolength{\footskip}{-0.8cm} 
As the non-paraxial optics is mathematically analogous to a quantum gravitational system with minimal length, 
we will first review the deformation of quantum mechanics produced by GUP 
\cite{KMM,Bagchi,Das}. This form of GUP  has shown to be very useful in various contexts 
\cite{Bhat,Dey_review} and, for a one-dimensional system it is given by
\begin{equation}\label{XPCOm}
[X,P]=i\hbar(1+3\beta P^2),
\end{equation}
with $\beta >0$ being a parameter having the dimension of the inverse squared momentum. 
In (\ref{XPCOm}), $X,P$ are the generalized observables, represented by $X=x, P=p(1+\beta p^2)$ \cite{Das}
up to first order in $\beta$ in terms of canonical variables $x,p$ satisfying $[x,p]=i\hbar$. While, this is one of the possible representations, one may refer to \cite{DFK} for several other representations of the commutation relation \myref{XPCOm}.
Nevertheless, following \myref{GUP}, the GUP corresponding to our system \myref{XPCOm} turns out to be of the form
\begin{equation}
\Delta X\Delta P\geq\frac{\hbar}{2}\big[1+3\beta(\Delta P)^2\big],
\end{equation}
with $\langle P\rangle=0$, so that the minimal length corresponding to it can 
be computed as $\Delta X_0\sim\hbar\sqrt{\beta}$. The time-dependent free particle Schr\"odinger equation 
corresponding to the given representation is given by 
\begin{equation} \label{GSE}
i\hbar\partial_t\Psi=\frac{p^2}{2m}\Psi+\frac{\beta}{m}p^4\Psi,
\end{equation}
which in the position representation takes the form
\begin{equation} \label{GSEz}
i\partial_t\Psi+\frac{\hbar}{2m}\partial^2_x\Psi-\frac{\hbar^3\beta}{m}\partial^4_x\Psi=0.
\end{equation}
Note that this equation (\ref{GSEz}) holds true in a one-dimensional planar optical waveguide with $p$ being the transverse momentum ($p=-i\hbar\partial_x$) in the position representation \cite{Conti2014,Braidotti2016}. Seemingly, Eq. (\ref{GSEz}) also holds in an optical fiber for which $p=-i\hbar\partial_t$ is the time evolution operator. However, in this case one needs to neglect the third order dispersion as discussed in detail in \cite{Braidotti2016}. Nevertheless, in order to obtain an equivalent equation for optical system, we write the Helmholtz equation
\begin{equation}\label{helmholtz}
\nabla^2 \mathcal{E}+k_0^2\mathcal{E}=0,
\end{equation}
where $k_0$ is the wave-vector, such that $|k_0|=2\pi/\lambda$ and $ \lambda$ is the wavelength, and $\mathcal{E}$ is the electric field. The $\nabla$ operator represents the Laplacian $(\partial_z,\partial_x)$. 
The solutions of Eq. \myref{helmholtz} are forward and backward propagating waves with wave-number 
\begin{equation} \label{disp_rel}
k_z=\pm\sqrt{k_0^2-k_x^2}.
\end{equation}
Retaining only the forward propagating wave $(k_z>0)$, the forward projected Helmholtz equation (FPHE) reads as
\begin{equation} \label{FPHE}
i\partial_z\mathcal{E}+\sqrt{\nabla^2+k_0^2}\mathcal{E}=0.
\end{equation}
Here, we have neglected the vectorial effects which are not present in the vacuum propagation
\cite{Ciattoni2005,Longhi2011,Aiello2005} but may include the presence of evanescent waves, which however we exclude by assuming $|k_x|<k_0$. 
Expanding $k_z$ for small $k_x$ up to the second order, we obtain the unidimensional normalized
higher-order propagation equation \cite{Braidotti2016}
\begin{equation} \label{hope}
i\partial_{z}\psi+\frac{1}{2}\partial^2_{x}\psi-\frac{\varepsilon}{8}\partial^4_{x}\psi=0.
\end{equation}
Here, $\psi$ is the envelop wave-function proportional to the electric field, $z$ 
is the propagation direction, $x$ represents either the spatial or temporal variable and $\varepsilon$ is a 
dimensionless parameter. This equation and the Generalized Schr\"{o}dinger's equation are linked by the mapping $(x,t) \leftrightarrow (x,z)$ and $\hbar \leftrightarrow 1/k_0$. We can also observe that the $z$ variable has an associated energy variable as its Fourier conjugate. In the spatial case $\varepsilon = 1/k_0Z_d$, with $Z_d$ being the diffraction length.
Whereas, in the temporal case $\varepsilon=-\beta_4/(3\beta_2T^2_{0})$, with $T_0$ being the initial temporal 
pulse duration and $\beta_2,\beta_4$ being the second and {  fourth} order dispersion coefficients, respectively. It is clear that the differential operator of equations \myref{GSEz} and~\myref{hope} are not bounded from below. As we shall see in the next section, this can be linked to the existence of inherited minimal length in the system.
It should be noted that the light propagation equation \myref{hope} has the same form as the GUP deformed Schr\"odinger 
equation \myref{GSE}, apart from a difference in parameter. This mathematical analogy allows to apply techniques coming 
from quantum gravity theories to analyze non-paraxial and ultrafast regimes 
for optical propagation and to associate the minimal measurable length scale with the optical resolution. Indeed, 
it has been proved recently \cite{Braidotti2016} that the generalized Schr\"{o}dinger equation is
a novel tool for describing short pulses and ultra-focused beams, which predict the existence of 
a minimal spatial or temporal scale. Besides, different schemes have been developed on the use of 
such a mathematical approach to optics; such as, in the analysis of wave propagation in random or 
gradient-index media \cite{gmezreino1987,Eichmann1971}.
\section{Path Integral}\label{sec:path}
Path integral method is a powerful mathematical to obtain the propagator of a system, and   
has been used in various physical systems, 
especially in quantum field theory and statistical mechanics \cite{Feynman1948}. 
A recent study demonstrated that this approach can be applied to find the kernel of several 
GUP deformed non-relativistic systems, particularly, the free particle \cite{Das1}. 
Our intent here is to obtain the path integral kernel in order to reveal that the presence of GUP causes 
various effects on non-paraxial and higher order optics. It is worth noting that even though 
the equation governing our optical system has the same form as that of quantum gravity, 
the physical meaning of the path integral is different from that of the GUP deformed Schr\"odinger equation. 
In the GUP deformed Schr\"odinger equation, the Feynman integral represents the sum of all
paths for a classical particle on a deformed background geometry. Whereas, in the optical system the path integral
represents the sum over all optical trajectories for a non-paraxial beam in the spatial case (or, for a pulse
in the temporal domain) \cite{Eichmann1971}. All the optical ray paths contribute to the propagator. 
However, only the one which adds coherently contributes to the final field. The difference between the
optical and the quantum path integrals is the unit of the phase term, which in the optical case is measured
in units of $k_0^{-1}$, instead of $\hbar$ as in quantum mechanics. 
We also consider the  effective time $ t= z/c$ 
for the beam, in order to make the similarities between GUP-deformed quantum mechanics and the
non-paraxial optics more prominent.

\begin{figure}[h!] \label{fig::prob}
		\centering
		\includegraphics[width =0.75 \textwidth]{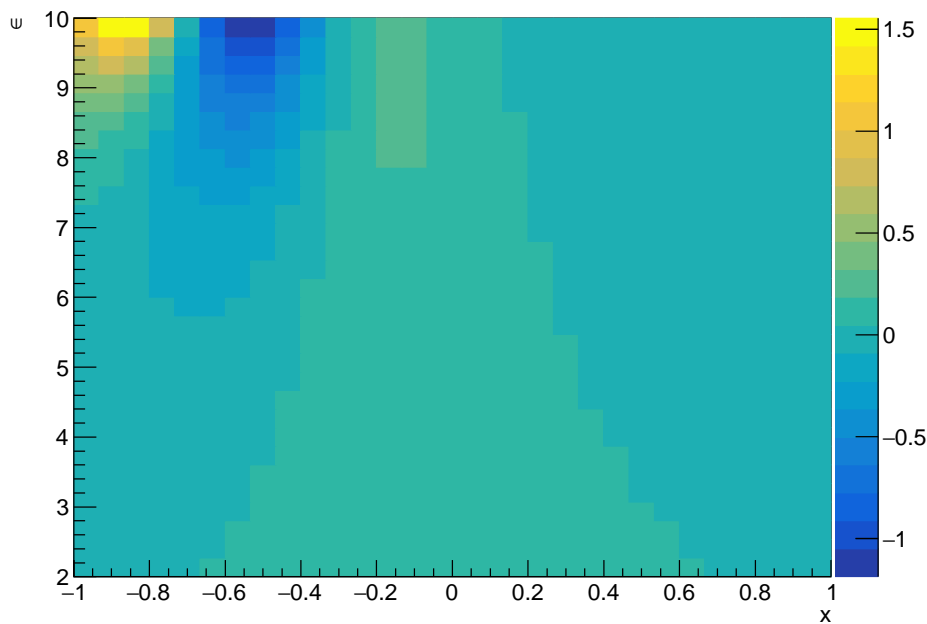}
		\caption{A normalized probability density solution~$|\psi(z,x)|^2$ with RMS $ \frac{1}{\sqrt{2}}$ as a function of the position along the $x$-axis and the parameter $\varepsilon$ along the $y$-axis. We observe that if the parameter $ \varepsilon > 3$, we get regions on which the probability density is negative making such configurations (of RMS and $\varepsilon$) unphysical. Otherwise, there would be a catastrophic production of pulses.  }
	\end{figure}

Let us consider a trajectory for a optical pulse going from $x'$ to $x''$ in a short effective time interval $\Delta t$, such that In the spatial case, we can write the kernel for the optical system as 
\begin{align}\label{2}
K(x'',t'+&\Delta t;x',t') =\int{e^{i k_x(x''-x')}e^{-i \left(\frac{k_x^2\Delta t}{2}+\frac{\varepsilon\Delta t}{8}k_x^4 \right)}}\frac{dk_x}{2\pi },
\end{align}
which when expanded in $\Delta t$, we find 
\begin{equation}\label{3}
K(x'',t'+\Delta t;x',t')=\delta (x''-x')-i\Delta t \left[-\frac{1}{2}\frac{\partial^2}{\partial {x''}^2}\delta (x''-x')+\frac{\varepsilon}{8}\frac{\partial^4}{\partial {x''}^4}\delta (x''-x')\right].
\end{equation}
We observe that in the limit of $\varepsilon \rightarrow 0$ we obtain the kernel of the standard path integral \cite{Claud}. However, we can rewrite Eq. \myref{2} as 
\begin{equation} \label{4}
K(x'',t'+\Delta t;x',t')=\int e^{i \int_{t'}^{t'+\Delta t}
\left(k_x\dot{x}-\frac{k_x^2}{2}-\frac{\varepsilon}{8}k_x^4\right)dt}\frac{dk_x}{2\pi },
\end{equation}
so that after a straightforward calculations we arrive at \cite{Das1}
\begin{align}\label{5}
&K(x'',t'+\Delta t;x',t')=\frac{1}{\sqrt{2\pi i \Delta t}}\left[1+\frac{3\varepsilon i }{8\Delta t}-\frac{3\varepsilon(x''-x')^2}{4\Delta t^2}-\frac{i\varepsilon(x''-x')^4}{8 \Delta t^3}\right] e^{\frac{i(x''-x')^2}{2 \Delta t}}.
\end{align}
Furthermore, for a  finite interval ($t''-t'=N\Delta t$) we have \cite{Das1}
\begin{align} \label{6}
&K(x'',t'';x',t')=\frac{1}{\sqrt{2\pi i (t''-t')}} e^{\frac{i(x''-x')^2}{2 (t''-t')}}\left[1+\frac{3\varepsilon i }{8(t''-t')} -\frac{3\varepsilon(x''-x')^2}{4(t''-t')^2}-\frac{i\varepsilon(x''-x')^4}{8 (t''-t')^3}\right]. 
\end{align}
This Kernel satisfies the equation \myref{hope} describing the higher order propagation equation for ultra-short pulses and ultra-focused beams. The solution to this higher order propagation equation can be written as \cite{Das}
\begin{align} \label{8}
\psi(x,z)&=Ae^{ik_x(1-\varepsilon  k_x^2/8)x-iEz }+Be^{-ik_x(1-\varepsilon k_x^2/8)x-iEz }+Ce^{\frac{2x}{\sqrt{\varepsilon}}-iEz }+De^{-\frac{2x}{\sqrt{\varepsilon}}-iEz },
\end{align}
which consists of a new exponential terms which are vanished in the limit $\varepsilon\rightarrow 0$. 
We also observe that the solution consists of a superposition between incoming and outgoing pulses, 
in terms dispersion direction $x$. Such that for an incoming wave,
we have $ A= C=0$, and for the outgoing one, we have  $ B=D=0$. Furthermore, it is easy to see that
the relations between these constants are $A=-C$, and $B=-D$. In quantum field theory, the kernel make the system $\psi(x,z)$ propagated from $(x',z')$ to $(x'',z'')$, so that
\begin{equation}
\psi(x'',z'')=\int K(x'',z'';x',z')\psi(x',z')dx'. 
\label{9}
\end{equation}
It is worth noting that although this formalism is generally used for quantum particles, however, since we have am equivalence between optical and quantum system, we can apply this result directly to the optical systems. In QFT, the probability that a particle arrives at the point $x''$ is proportional to the square modulus of the propagator $K$, $|K(x'',x',z''-z')|^2$. Analogously, $|K(x'',x',z''-z')|^2$ gives the intensity of the optical beam at the point $x''$ 
\begin{align}\label{12}
P(x'')dx&=K^*(x'',z'',z''-z')K(x'',z',z''-z')dx=\left(1-\frac{3\varepsilon(x''-x')^2}{2(z''-z')^2}\right)\frac{1}{2\pi (z''-z')}dx.
\end{align}
Thus, it is obvious that the path integral formalism sheds light on the wave-particle duality of light. Since, the propagator $K$ is proportional to the field amplitude $\psi$, $|K|^2$ gives the probability of having a certain value of the field in a certain point of space. This probabilistic view is a direct consequence of diffraction of light. Furthermore, as $\varepsilon>0$, the term $1-\frac{3\varepsilon(x''-x')^2}{2(z''-z')^2}$ is smaller than one. However, since the probabilities have to be positive definite, we obtain a restriction on the system
\begin{equation}
1-\frac{3\varepsilon(x''-x')^2}{2(z''-z')^2}\geq0,
\end{equation}
which, in turn, provide the bound on the momentum for the analogous optical system 
\begin{equation} \label{13}
p\leq p_{max}=\sqrt{\frac{2}{3\varepsilon}}. 
\end{equation}
The momentum of the pulse is linked to its wavelength by~$p = 2\pi/k_0 \lambda $ and, hence, \myref{13} implies a minimal wavelength of the pulse given by
	\begin{equation}
\lambda_{\text{min}} = \sqrt{\frac{6 \pi^2}{k_0^3 Z_d}}.
	\end{equation}
	The fact that the equations \myref{GSEz} and~\myref{hope} are unbounded from below can be interpreted by the negative probabilities resulting from pulses of wavelength shorter than $\lambda_{\text{min}}$ which are unphysical.  This could also be seen when trying to squeeze the pulse of the beam with root-mean-square (RMS) smaller than $1/2p_{\text{max}}$. This results in an unphysical behavior for the 
	wave-packet as seen in Fig.~\ref{fig::prob}.
	  These results are in perfect agreement with the theory of diffraction of light and, thus,
it reveals the power of the path integral formalism with respect to the standard Fourier analysis.
It may also allow us {to reveal} many important features in a much easier way. It is obvious that the
presence of a minimal length scale and, hence, a minimal spatial extension of the beam waist affects the 
transverse spectral distribution of the optical modes. This means that the optical beam considered here
can access only a limited spectral band given by Eq. \myref{13}. 

It may be noted that for  analyzing any optical (or any system involving electromagnetic radiation), 
we should have use second quantized quantum electrodynamics. However, from the perspective of effective field theory, as 
we are dealing with scales much larger than the scale of an electron, we can neglect the loop corrections. So, this system 
can be analyzed as a first quantized theory. Now a first quantized field theory is mathematically analogous to the wave 
mechanics of a particle, and hence we are justified to use the mathematical  analog of a wave mechanics to analyze this 
system.  Thus, we find the mathematically  a non-paraxial 
optical system is  analogous to a GUP deformed quantum mechanical system.  We use this analog to justify the use of 
path integral of a GUP deformed quantum system  as a mathematical tool to analyze a non-paraxial optical system. 

\section{Instanton and Berry  phase}\label{sec:inst}

  As we have the path integral for 
such a optical system, we can study non-trivial effects for such a system. In fact, we do use this 
formalism to calculate the effects of instanton and Berry phase for such an optical system. It may be noted 
that by doing that we demonstrate that  the Gouy phase (which is known to be a manifestation of 
a general Berry phase that occurs through a beam focus when the beam parameters undergo cyclic motion 
\cite{Subbarao1995,Winful2001}), in a non-paraxial system is trivial. This was not something which was obvious, 
and this demonstrates a useful application of the path integral formalism in optics. However, 
the main motivation to define the path integral would be to demonstrate that such non-trivial non-perturbative
effects can be analyzed using this formalism. 

Instantons were first introduced in Yang-Mills theory, which are the classical solutions of
equations of motion in the Euclidean space with nontrivial topology \cite{Grafke2015,Belavin1975}.
The instanton approach can be used as a semi-classical non-perturbative approach to find path integrals solutions. 
Instantons are saddle point configurations of the path integrals being equivalent to minimizers of the
related Euclidean action\cite{Coleman1979,Zakharov1982}. These solutions are usually able to characterize rare 
events in physical systems described by path integrals, like, strong vortex tubes in fluid flows, current sheets 
in plasmas, shocks in high Mach number flows \cite{Eyink2006,Yakhot1986}. The main mathematical tool used to calculate 
the instantons is the path integral, and as we have the path integral for non-paraxial optics, we can calculate the 
instantons for such a system. 
 
So, now we calculate instantons for a  non-paraxial optical system. 
Therefore, the calculus of instantons in optical systems corresponds closely to the treatment of instantons
in quantum field theory. The path integral can be written in an alternative form via the Legendre transformation
\begin{equation}
K(x'',t'';x',t') = \int \mathcal D [x] e^{i S/\lambdabar},
\label{path}
\end{equation}
where $S$ is the classical action having the form
\begin{equation}
S = \int  \left(  \frac{1}{2} \dot x ^2 - \frac{\varepsilon}{8} \dot x^4\right)  dt.
\end{equation}
In order to compute the instanton solution, it is necessary to preform a
Wick rotation $t \rightarrow i \tau$, which gives the Euclidean path integral 
\begin{equation} \label{Epath}
K_E(x'',\tau'';x',\tau') = \int \mathcal D [x] e^{ S_E},
\end{equation}
where  $S_E$ is the Euclidean action of the form
\begin{equation}
S_E = -\int   \left( \frac{1}{2} \dot x ^2 + \frac{\varepsilon}{8} \dot x^4 \right) d\tau.
\end{equation}
By varying this action with respect to $x$, we obtain the instanton equation of motion
\begin{equation}
\frac{d}{d \tau} \left[  \frac{d x}{d \tau} + \frac{\varepsilon}{2}\left( \frac{d x}{d \tau}\right) ^3 \right] =0,
\end{equation}
The solution to this equation can be written in the form    
\begin{equation}
\left( \frac{d x}{d\tau}\right)  = i\sqrt{\frac{2}{3 \varepsilon}}.
\end{equation}
It may be noted, that when rotated back to $ t$,  we obtain
\begin{equation}
p  = \sqrt{\frac{2}{3 \varepsilon}}.
\end{equation}
Interestingly, it matches exactly with the maximum momentum that we have obtained 
in the previous section in Eq. \myref{13}. Nevertheless, the instanton equation of motion can also be written as follows
\begin{equation}
x = \frac{1}{c} \sqrt{\frac{2}{3 \varepsilon}} \, z \,,
\end{equation}
indicating the condition for extrema in $ S_E$. These instanton solutions correspond to the 
extremal conditions of propagation of non-paraxial beams and ultrashort pulses, 
that is the maximal localization conditions previously found. 

The Berry phase is a geometric or topological phase acquired by a system after it is cycled though
a closed loop in parameter space. In optics, the Gouy phase or geometrical phase is known to be a manifestation of 
a general Berry phase that occurs through a beam focus when the beam parameters undergo cyclic motion 
\cite{Subbarao1995,Winful2001}. Here, we compute the geometric phase for the pulse using 
the path integral \myref{path}. Taking a closed loop over the $x$-space \cite{Anandan:1997kh,haken}, we find
\begin{equation}
\,^{x} \theta_B(k) = \frac{\oint_\gamma \mathcal D[x]\, \hat{ p}_0  e^{iS/\lambdabar}} {\oint_\gamma \mathcal D[x] e^{iS/\lambdabar}} = i \oint_\gamma \psi^\dagger_k(x,z) \frac{\partial}{\partial x}\psi_k(x,z) dx,
\end{equation}
where $ \psi_k(x,z)$ is the solution to the Schr\"{o}dinger-like equation \myref{hope} for a particular $k$. 
Substituting the wave function in the above expression, we obtain 
\begin{equation}
\,^{x}\theta_B(k) = i \oint_\gamma dx \left[  \left( |A|^2- |B|^2\right)  ik \left(1-\frac{ \varepsilon k^2}{8}\right) + \frac{2}{\sqrt{\varepsilon}} \left( |C|^2-|D|^2\right) \right].
\end{equation}
In fact, the value of this integral is equivalent to the expected value of the momentum $ \langle p \rangle$ 
taken with periodic boundary conditions, as 
the coordinate $x$ correspond to the direction of dispersion. In fact, it is evident that  the periodic boundary 
condition
\begin{equation}
\psi(x,z) = \psi(2\pi L+x,z)
\end{equation}   
would imply that the wave function is real, and hence the expected value of the momentum would vanish.
Making the Berry phase trivial for cycles along the $ x$ direction.
Taking $z$ as the cyclic variable, the Berry phase can be written as
\begin{equation}
\,^{z}  \theta_B =  \oint_C dz \left( |A|^2+|B|^2+|C|^2+|D|^2\right)  E.
\end{equation}
Thus, we obtain the following result for Berry of this optical system along $z$ direction
\begin{equation}
\,^{z} \theta_B = \alpha E T,
\end{equation}
where $T$ is the period of pulse, and $ \alpha$ is a  dimensionless constant depending on the normalization.
Since, $ E = n\hbar\omega$ and $ \omega = 2 \pi/T$, 
the Berry phase for such system, is given by
  \begin{equation}
  \,^{z} \theta_B = 2 \pi\alpha \hbar (n) \;\;\;\;\; n = 0,1,2,\dots
  \end{equation}
  Hence, Berry phase for the direction of propagation of the pulse is also trivial.  
  Implying that the topology of the Hamiltonian phase space remain trivial even 
  when the such non-paraxial quantum optical systems. However, it would be interesting to analyze 
  the Berry phase for such systems  with non-linear terms, and use this formalism to obtain the Berry 
  phase for such non-linear optical systems. The important result in this paper is that a formalism to 
  calculate the Berry phase for non-paraxial optical systems was provided, and this formalism 
  can now be used to calculate the Berry phase for other optical systems. 
\section{Conclusion}\label{sec5}
In this paper, we have used the mathematical analogy between a quantum gravitational system with minimal length and 
a non-paraxial quantum optical system to obtain a path integral for such a non-paraxial optical system.
This is possible because  
  mathematically such an optical system resembles a deformed  
  Schr\"odinger equation in a quantum gravity, where the deformation has occurred  due to the existence
 of an intrinsic minimal length in spacetime. It may be noted that the reason behind this mathematical analogy 
 is the presence of such a minimal measurable length scale in non-paraxial optical system. In fact, we show that 
 this optical system contains an upper bound for the momentum which occurs due to the   $\beta$ corrections.
We have constructed the path integral using this mathematical analogy, 
 and verified the consistency properties thoroughly.  We have also used such a path integral to calculate instantons in 
 this optical system. After calculating the instantons for this optical system, we have also studied the Berry 
 phase of this optical system using this path integral. Thus, we explicitly calculated the Berry phase along both 
 $x$ and $z$ directions. It may be noted that the formalism used to obtain to obtain the Berry phase for non-paraxial 
 optical system can be used to obtain the Berry phase for other optical systems. It would be interesting to investigate 
 the Berry phase for a non-linear optical system, using this formalism. This analysis can also  be 
extended in several directions, such as, retaining higher order dispersion. Furthermore, 
our results show that quantum mechanics deformed by   an intrinsic minimal length in spacetime
is mathematically similar to a non-paraxial optical system. As this minimal length in spacetime can be motivated from 
quantum gravity, the paraxial quantum optics can be used as an quantum gravity analog in optics. It may be noted that 
 various optical   analogues quantum gravitational  have been studied 
\cite{Faccio2012, Barcel2003, Longhi2011, Parades2016}.  It would be interesting  to use the path integral obtained in 
this paper, to analyze such systems. In fact, it would be interesting to investigate the Berry phase for such systems, 
using the formalism developed in this paper. 

\vspace{0.5cm}
\noindent \textbf{\large{Acknowledgements:}} C.C. acknowledges support from the Templeton foundation (Grant number 58277). S.D. is supported by an INSPIRE Faculty grant (DST/INSPIRE/04/2016/ 001391) by the Department of Science and Technology, Govt. of India. S.A. was supported by a grant from the “Research Center of the Female Scientific and Medical Colleges”, Deanship of Scientific Research, King Saud University.



\end{document}